\documentclass{article}
\usepackage{graphicx} % Required for inserting images
\usepackage{appendix}

\usepackage[hypertexnames=false]{hyperref}
\hypersetup{
    colorlinks=true,
    linkcolor=blue,
    filecolor=magenta,   
    citecolor=black,   
    urlcolor=cyan,
    }

\def\red#1{\textcolor{red}{Jarah: #1}}
\def\blu#1{\textcolor{blue}{Emilio: #1}}

\begin{document}
\def\thefootnote{\fnsymbol{footnote}}
\def\thetitle{Finite Energy Resolution, Correlations Between Bins and Non-Nested Hypotheses}
\def\auttwo{Jarah Evslin}
\def\autone{Emilio Ciuffoli}
\def\affa{Institute of Modern Physics, NanChangLu 509, Lanzhou 730000, China}
\def\affb{University of the Chinese Academy of Sciences, YuQuanLu 19A, Beijing 100049, China}

\title{Titolo}

\begin{center}
{\large {\bf \thetitle}}

\bigskip

\bigskip

%\catcode`@=11

{\large \noindent  \autone{${}^{1}$}\footnote{emilio@impcas.ac.cn} and \auttwo{${}^{1,2}$}}\footnote{jarah@impcas.ac.cn}

%{\large \noindent  \autone{${}^{1,2}$} \footnote{jarah@impcas.ac.cn} and \auttwo{${}^{1,2}$} \footnote{guohengyuan@impcas.ac.cn}}

\vskip.7cm

1) \affa\\
2) \affb\\
%3) \affc\\

\end{center}

\begin{abstract}
    We show that whenever independent events are binned, there is no correlation between bins.  In particular, neither the uncertainty in the energy resolution nor a systematic error in the energy response of a detector will lead to a correlation between bins.  This is in contrast with a result that recently appeared in the literature, where it was claimed that the finite energy resolution could induce correlation between bins, changing the distribution of $\Delta\chi^2$ when attempting to determine the neutrino mass ordering using reactor neutrinos. We will compute the expression for the variance of $\Delta\chi^2$ in the case of non-nested hypotheses if correlations between bins are present, showing that they could indeed affect the distribution. We will also show, however, that the detector response cannot introduce statistical correlations between bins unless these are already present in the original spectrum. Both of these results are valid in general: the absence of correlation is true for any detector response and the distribution of $\Delta\chi^2$ in the presence of (intrinsic) correlations, while obtained having the mass ordering in mind, it is true for any non-nested hypotheses.
\end{abstract}

\setcounter{footnote}{0}
\renewcommand{\thefootnote}{\arabic{footnote}}

\section{Introduction}

During the next years, several experiments will attempt to robustly determine the neutrino mass ordering, such as JUNO (reactor neutrinos)~\cite{JUNO:2015zny}, Hyper-Kamiokande~\cite{Hyper-Kamiokande:2018ofw} and DUNE~\cite{DUNE:2020ypp} (accelerator neutrinos). Atmospheric neutrino experiments, such as Super-Kamiokande~\cite{Super-Kamiokande:1998kpq} and IceCube~\cite{IceCube:2014flw}, have some sensitivity to the mass hiearachy as well. There are already some preliminary results, mostly using the data from T2K~\cite{T2K:2011qtm} and NO$\nu$A~\cite{NOvA:2004blv}, that provide a hint for the normal ordering \cite{Esteban:2020cvm}.

10 years ago, it was pointed out that, since the two hierarchies are non-nested hypotheses, Wilks' theorem cannot be applied and $\Delta\chi^2$ follows a Gaussian distribution, with average $\overline{\Delta\chi^2}$ and standard deviation $\sigma=2\sqrt{\overline{\Delta\chi^2}}$\footnote{here and in the rest of the paper, a bar over the variable will be used to indicate its expected value, e.g. $\overline{x}$ will indicate the expected value of $x$}\cite{Qian:2012zn,Ciuffoli:2013rza,Blennow:2013oma}. 

This conclusion was verified using Monte Carlo (MC) simulations.  One of the issues faced in those computations is that the number of events required to simulate each experiment can be very high (for example, for reactor neutrino experiments, such as JUNO, it is of the order of $10^5$). A solution commonly adopted is to compute the expected number of events for each bin $\overline{N_i}$; the number of events in a given bin for a particular experiment will be given by 
\[
N_i=\overline{N_i}+g_i\sqrt{\overline{N_i}}
\]
where $g_i$ is a Gaussian variable with average 0 and standard deviation 1\footnote{Technically, the fluctuations follow a Poisson distribution, but since $\overline{N}_i\gg1$, a Gaussian distribution is an excellent approximation}.

Recently, however, it was emphasized~\cite{Sawy:2023eyu} that the finite energy resolution will modify the distribution of $\Delta\chi^2$, in particular for reactor neutrino experiments, such as JUNO. The authors noted that, if the data are created on an event-by-event basis, first generating the "real" spectrum, {\it i.e.} the one that would be observed with perfect energy resolution, then applying the energy smearing later, the distribution of $\Delta\chi^2$ is considerably wider, which leads to a lower sensitivity to the mass ordering. 

In this paper, first we will compute the theoretical distribution of $\Delta\chi^2$ for two non-nested hypotheses if the correlations between bins are not negligible, showing that they can, indeed, affect the variance. Then, we will calculate the expected value of the correlations between bins, assuming finite energy resolution. Our results, however, are in tension with what is reported in Ref.~\cite{Sawy:2023eyu}: indeed we will prove that the detector response (regardless of whether it arises from energy resolution or any other effect) cannot introduce correlations unless they are already present in the original spectrum (which is not the case for reactor neutrinos). We will see, indeed, that the finite energy resolution leads to two sources of fluctuations that exactly cancel each other in the non-diagonal elements of the covariance matrix: the ones that follow a Poisson distribution, present in the original spectrum, and the ones that follow a multinomial distribution, due to the variation on the number of events that migrate from one bin to another. We stress that, while these results were obtained in the context of the neutrino mass ordering determination, they are quite general: the absence of correlations related to the detector response is not just valid only for neutrino experiments, or limited only to energy resolution. Moreover, the formula obtained regarding the distribution of $\Delta\chi^2$ in the presence of correlations between bins (that could be present in the original spectrum) is true for any non-nested hypotheses. 

This paper is structured as follows.  In Sec.~\ref{sec::Chi}, we will generalize the results obtained in Ref.~\cite{Qian:2012zn} and \cite{Ciuffoli:2013rza}, calculating the distribution of $\Delta\chi^2$ for non-nested hypotheses in the presence of correlations between bins, both in the case Simple vs. Simple ({\it i.e.} no pull parameters are taken into account), as well as when pull parameters are present. We will also discuss what happens if the dependence of the spectrum on the pull parameter(s) is significantly different for the two hypotheses.

In Sec.~\ref{sec::Correlation} we will first present a thought experiment, to give an intuitive understanding of the issue, then we will compute the correlation between bins when there is a smearing due to the detector response.

In Sec.~\ref{sec::Toy}, we will check the results obtained previously using two toy models: a very simplified one (Very Simple Toy Model, or VSTM), which will allow us to us large MC samples without requiring too much computation time, and one more similar to JUNO (JUNO Toy Model, or JTM).

\section{Distribution Of $\Delta\chi^2$ and Correlations}
\label{sec::Chi}

\subsection{Simple vs. Simple}
%Let us assume there are two hypotheses, $N$ and $I$ (in the case of the mass ordering, they would represent normal and inverted order, respectively; however this argument is valid in general when testing non-nested hypotheses). 
%We will call $\overline{N}_{H,i}$ the expected number of events in the i-th bin under the hypothesis $H$ (with $H=N,I$). In general, the number of events in a particular experiment if the hypothesis $H$ is true can be expressed as
%\begin{equation}\label{eq::DefN_gen}
%    N_{H,i}=\overline{N}_{H,i}+\delta N_{H}^I
%\end{equation}
%For now, we are not making any assumption on $\delta N_{H}^i$, except that, by definition, its expected value must be zero, {\it i.e} using $\langle ... \rangle$ to indicate the average over all the experiments, we have
%\[
%\langle N_{H,i} \rangle \equiv \overline{N}_{H,i} \qquad \Rightarrow \langle \delta N_{H}^i\rangle=0
%\]

There are many situations in which the event counts in different bins will be correlated.  For example, if the strength of the source is unknown, then a larger count in one bin suggests a stronger source which suggests a large event count in all bins.  Thus one expects a positive correlation.  Conversely, if the total number of events is fixed, then more events in one bin leaves less for the others, and so one arrives at a negative correlation.  

We will now derive the distribution of $\Delta\chi^2$ for non-nested hypotheses if the covariance matrix has non-diagonal elements. The origin of these correlations is irrelevant for our purposes, they could be intrinsically present in the spectrum, or due to other factors. Later we will focus in particular on the correlations due to the finite resolution of the detector.

For simplicity, first we will consider a Simple vs. Simple case, {\it i.e.} no pull parameters will be taken into account; then we will show that the presence of pull parameters (under certain assumptions) does not change significantly the result. %Their inclusion, however, is quite straightforward (see Ref.~\cite{Ciuffoli:2013rza}) and will not affect the results. 

Let us assume there are two hypotheses, $N$ and $I$ (in the case of the mass ordering, they would represent normal and inverted order, respectively; however this argument is valid in general when testing non-nested hypotheses). Let us define
\begin{equation}
    \langle x \rangle =\overline{x}
\end{equation}
to be the average value of the generic variable $x$. We will call $\overline{N}^O_{H,i}$ the expected number of events in the $i$-th bin if the hypothesis $H$ (with $H=N,I$) is true. We will consider the case in which, in Nature, the hypothesis $N$ is true.  The number of events observed in an actual experiment can be written as
\begin{equation}
    N^O_{i}=\overline{N}^O_{N,i}+\delta N_i.
\end{equation}
Please note that we do not need any particular assumption regarding $\delta N_i$, except that, by construction, $\langle \delta N_i\rangle=0$. For each hypothesis, $\chi_H^2$ will be defined as
\begin{equation}\label{eq::defChiGen}
\chi^2_H=\sum_i\frac{(N^O_{i}-\overline{N}^O_{H,i})^2}{\sigma_i^2}
\end{equation}
where $\sigma_i^2=\langle (\delta N_i)^2\rangle$%\footnote{Using a Bayesian approach, one should use $\sigma_{H,i}^2=\langle (\delta N_{H}^i)^2\rangle$, based on which hypothesis is used; in other approaches people assume $\sigma_{H,i}^2=N^O_i$. For the determination of the mass hierarchy using  reactor neutrinos, the difference between those approaches is negligible, since $\langle (\delta N_{H}^i)^2\rangle=\overline{N}^O_{H,i}$ and $\overline{N}^O_{N,i}\sim\overline{N}^O_{I,i}\gg1$} \blu{ Vale la pena di mettere una discussion (come quella che ho aggiunto nella nota) sul fatto che, in principio, see si fitta la gerarchia inversa uno dovrebbe usare $\sigma_i^2=\langle (\delta N_{I}^i)^2\rangle$ usando un approccio Bayesiano, oppure il numero di eventi osservati (see si usa un approccio frequetista), ma la differenza e` minima se $N_N\gg1$ e puo` essere trascurata?}.

We have
\begin{equation}
\label{eq::defChi}
\chi^2_N=\sum_i\frac{(\delta N_i)^2}{\sigma_i^2}, \qquad \chi^2_I=\sum_i\frac{(\Delta N_i+\delta N_i)^2}{\sigma_i^2}
\end{equation}
where $\Delta N_i=\overline{N}^O_{N,i}-\overline{N}^O_{I,i}$.

$\Delta\chi^2$ is defined to be
\begin{equation}\label{eq::DeltaChiCasoSemplice}
\Delta\chi^2=\chi^2_I-\chi^2_N=\overline{\Delta\chi^2}+\sum_i\frac{\Delta N_i\delta N_i}{\sigma_i^2}, \qquad \overline{\Delta\chi^2}=\langle\Delta\chi^2\rangle=\sum_i\frac{\Delta N_i}{\sigma_i^2}.
\end{equation}
The variance its distribution is
\begin{equation}\label{eq::varChi}
\textrm{Var}({\Delta\chi^2})=\langle (\Delta\chi^2-\overline{\Delta\chi^2})^2\rangle=4\sum_{i,j}\frac{\Delta N_i\Delta N_j}{\sigma_i^2\sigma_j^2}\langle\delta N_i \delta N_j\rangle.
\end{equation}

If we assume, as it is done in \cite{Qian:2012zn} and \cite{Ciuffoli:2013rza}, that $\delta N_{N}^i$ are Gaussian fluctuations, uncorrelated one to another, we would obtain the usual formula
\begin{equation}\label{eq::CasoGaussiano}
\langle\delta N_i \delta N_j\rangle=\delta_{ij}\sigma_{i}^2 \qquad \textrm{Var}({\Delta\chi^2})=4\overline{\Delta\chi^2}.
\end{equation}

\subsection{Pull Parameters}\label{sec:PullParameters}
Even if pull parameters are present, the distribution of $\Delta\chi^2$ does not change significantly, as long as some conditions are fulfilled.

%Let us first simply a bit the notation: since in this section we will not care about real/observed spectra, we will drop the index $R/O$. 
Let $x$ be the pull parameter. $x_{0,N(I)}$ will be the value of $x$ that minimizes the Asimov $\chi^2$ under the hypothesis $N$ $(I)$, $\Delta N_i=\overline{N}^O_{N,i}(x_{0,N})-\overline{N}^O_{I,i}(x_{0,I})$. 

In this subsection, it will be useful to switch to a vector notation: using $v^i$ to indicate the $i$-th component of the vector $\vec{v}$, we can define
\begin{equation}
    v_{\Delta}^i=\frac{\Delta N_i}{\sigma_i}, \qquad v_g^i=\frac{\delta N_{i}}{\sigma_i}, \qquad v_{H}^i=\frac{1}{\sigma_i}\left. \frac{\partial N^O_{H,i}(x)}{\partial x}\right|_{x=x_{0,H}}.
\end{equation}
The scalar product between two vectors $\vec{v}$ and $\vec{w}$ is defined, as usual, as
\begin{equation}
    \vec{v}\cdot \vec{w}=\sum_{i=1}^n v^iw^i
\end{equation}
%We will also use $|\vec{v}|=\sqrt{\vec{v}\cdot\vec{v}}$ \blu{Questo e` necessario specificarlo? Mi sembra superfluo, sinceramente...  }.

Eq.~(\ref{eq::defChi}) now reads
\begin{equation}\label{eq::defChiVector}
    \chi^2_N=(\vec{v}_g-\vec{v}_{N}\delta x_N)^2, \qquad \chi^2_I=(\vec{v}_\Delta+\vec{v}_g-\vec{v}_{I}\delta x_I)^2
\end{equation}
where $\delta x_H=x-x_{0,H}$. Since we are assuming that, in Nature, the hypothesis $N$ is true, $x_{0,N}=\overline{x}$, {\it i.e.} asymptotically the best fit value of $x$ obtained under the hypothesis $N$ will converge to the true value of $x$; this is not necessarily true for $I$. Let us assume that
\begin{equation}
\langle v_g^i v_g^j\rangle = \delta_{ij}.
\end{equation}
For reactor neutrinos, this would be trivially true with perfect energy resolution. In the next section, we will prove that this does not change even if the detector response is taken into account.

By requiring that $\chi_I^2$ computed with the Asimov dataset ({\it i.e.} $\vec{v}_g=0$) is minimized by $\delta x_H=0$, we have
\begin{equation}\label{eq::perpendicularityVx}
    \vec{v}_\Delta \cdot \vec{v}_{I}=0
\end{equation}
(such a condition is automatically verified for $\chi^2_N$, by construction). For a specific experiment, with a given value of $\vec{v}_g$, the value of $\delta x_H$ required to minimize $\chi^2$ is given by
\begin{equation}\label{eq::defDeltaX}
    \delta x_H=\frac{\vec{v}_g\cdot\vec{v}_{H}}{|\vec{v}_{H}|^2}.
\end{equation}
This is equivalent to projecting the vector $\vec{v}_g$ into the plane perpendicular to $\vec{v}_{H}$. 

%It is also useful to notice that the $\sigma_{H,x}$, which is the precision with which $x$ can be measured under the assumption $H$ and is defined as 
%\begin{equation}
 %   \chi_H^2(\delta x_H+\sigma_{H,x})-\chi_H^2(\delta x_H)=1
%\end{equation} 
%is given by
%\begin{equation}
%    \sigma_{H,x}=\frac{1}{|\vec{v}_{H,x}|^2}
%\end{equation}

It was shown in Ref.~\cite{Ciuffoli:2013rza} that a crucial assumption required for the distribution of $\Delta\chi^2$ to follow a Gaussian distribution and to satisfy Eq.~(\ref{eq::CasoGaussiano}) is that the two planes, for normal and inverted ordering, must be parallel.  This is the codition 
\begin{equation}\label{eq::ConditionParallel}
    \vec{v}_{N}=\vec{v}_{I}=\vec{v}_x \qquad \Rightarrow \delta x_{N}=\delta x_I=\delta x.
\end{equation}
Please note that the best-fit value for $x$ can still be different under the two hypotheses, namely $x_{0,N}\neq x_{0,I}$, it is just $\delta x$, {\it i.e.} the shift from these values that minimizes $\chi^2$ for a given experiment, that will be the same for both hypotheses. 

Under this assumption, Eq.~(\ref{eq::DeltaChiCasoSemplice}) can be rewritten as
\begin{equation}\label{eq::DeltaChiVettoriale}
    \Delta\chi^2=\overline{\Delta\chi^2}+2\vec{v}_\Delta\cdot(\vec{v}_g-\vec{v}_x\delta x) \qquad \overline{\Delta\chi^2}=|\vec{v}_\Delta|^2
\end{equation}
and Eq.~(\ref{eq::varChi}) becomes
\begin{equation}
    \textrm{Var}(\Delta\chi^2)=4\sum_{i,j}v_{\Delta}^iv_\Delta^j
    \langle (v_g^i-v_x^i\delta x)(v_g^j-v_x^j\delta x)\rangle.
\end{equation}
Using Eq.~(\ref{eq::defDeltaX}) we have
\begin{equation}
     \langle (v_g^i-v_x^i\delta x)(v_g^j-v_x^j\delta x)\rangle=\delta_{ij}-\frac{v_x^iv_x^j}{|\vec{v}_x|^2}.
\end{equation}
However, due to Eq.~(\ref{eq::perpendicularityVx}), the contribution from the second term will vanish and the variance of $\Delta \chi^2$ will take the usual form
\begin{equation}\label{eq::VarDeltaChi}
    \textrm{Var}(\Delta\chi^2)=4\overline{\Delta\chi^2}.
\end{equation}

If Eq.~(\ref{eq::ConditionParallel}) is not satisfied, we would still have, as in Eq.~(\ref{eq::DeltaChiVettoriale})
\begin{equation}
    \langle\Delta\chi^2\rangle=\overline{\Delta\chi^2}=|\vec{v}_\Delta|^2.
\end{equation}
The variance of $\Delta\chi^2$, however, now is
\begin{equation}
    \textrm{Var}(\Delta\chi^2)=4(\overline{\Delta\chi^2}+\textrm{Sin}^2(\alpha))
\end{equation}
where
\begin{equation}
    \textrm{Cos}(\alpha)=\frac{\vec{v}_{N}\cdot\vec{v}_{I}}{|\vec{v}_{N}||\vec{v}_{I}|}.
\end{equation}

\section{Correlation Between Bins}
\label{sec::Correlation}

We will now argue that smearing effects such as the finite energy resolution do not lead to a correlation between bins.  

We will present two arguments.  The first is a purely formal thought experiment, which can give an intuitive understanding of the issue.  The second is an analytic calculation showing that any effect that would cause migrations of events cannot introduce correlations if they are not already present in the spectrum.

\subsection{Thought Experiment}

Consider two experiments.  In the first, events in each energy range $i$ occur independently and at a rate $\lambda_i$.  The experiment runs for a unit of time.   In the detector, a given event will be assigned to precisely one bin $j$.  The probability that an event in energy range $i$ is assigned to bin $j$ is $G_{ij}$.  The matrix $G_{ij}$ is a proxy for the energy resolution.

In the second experiment,  the events in each energy range $j$ occur independently and at a rate $\sum_i G_{ij}\lambda_i$.  If the event occurs in the energy range $j$, it is automatically assigned to the bin $j$.  

We claim that in the case of both experiments, the distribution of events in the bin $j$ is a Poisson distribution with expectation value $\sum_i G_{ij}\lambda_i$ and that the covariance matrix between the number of counts per bin is diagonal, so there is no correlation between the bins. 

In the case of experiment two, this claim is trivial.  We defined the events to occur independently in each bin, and it is known that this condition leads uniquely to the Poisson distribution for each bin.

What about experiment one?  We claim that the two experiments are equivalent in the following sense.  Whatever process transmutes the energy $i$ into the bin $j$ in the detector in experiment one, we define to be part of the source and not part of the detector.  Now since the events were independent before this process, the events must be independent after as each event is treated independently.  Therefore, with this renaming, we have shown that experiment one and two are equivalent, and so since our claim was correct for experiment two it must also follow for experiment one.

Let us describe how this sleight of hand, redefining experiment one to experiment two, may be achieved in an example.  Consider a particle of energy $E$ inside of a scintillator.  Let us say that $E$ lies in the $i$th interval and so we assign it to bin $i$.  It has some chance $G_{ij}$ of creating $j$ photons that arrive at a given set of perfect PMTs.  The particles themselves are assumed to be created by an independent process, for example via the radioactive decay of a macroscopic source.  The scintillator is usually considered to be part of the detector, and so this is experiment one.  But one could equally well consider the PMTs to be the detector and the scintillator to be part of the experiment.  This is experiment two.  Clearly they are equivalent.  
Considered as experiment two, the energy resolution is perfect.  Therefore any correlation between the bins cannot result from the energy resolution.  

This highlights the fact that there is no inherent difference between the energy resolution of the detector and other smearings that arise at the source, such as the distributions of the energies in the various decays and thermal line broadening.  Therefore the energy resolution cannot lead to any additional uncertainties beyond that which one finds the traditional way, by considering the source with energy resolution smearing included.  Indeed, this is how most studies of JUNO have been performed.

%, but could only arise from the source, which generates events independently by assumption.

\subsection{Analytical Calculations}

Let us now check the above results directly, by calculating the covariance of the count numbers in the bins.  %Although we have argued that the covariance matrix between the bins will be exactly diagonal, in this subsection we will only consider the leading order expansion given a large number of events per bin.  This is already sufficiently precise to test the results of Ref.~\cite{Sawy:2023eyu}.

We will call $\overline{N}^R_{H,i}$ the expected number of events in the $i$-th bin with perfect energy resolution under the assumption $H$. Considering the finite energy resolution, %(and assuming that the dimension of the bins is sufficiently small, so we can use the continuum limit), 
the number of events we expect to observe will be
\begin{equation}
\overline{N}^O_{M,i}=\sum_jG_{ij}\overline{N}^{R}_{M,j}
\end{equation}
where the matrix $G_{ij}$ is the fraction of events that migrate from bin $j$ to bin $i$.

Let us define $E_{R(O)}$ to be the real (observed) energy of an event; the i-th bin goes from $E_{R(O),i}$ to $E_{R(O),i+1}$. Let us also define $\phi(E_R)$ to be the energy distribution of the events and $G(E_R,E_O)$ to be the response of the detector, we have
\begin{equation}\label{eq::defGgen}
G_{ij}=\frac{\int_{E_{O,i}}^{E_{O,i+1}}\textrm{d}E_{O}\int_{E_{R,j}}^{E_{R,j+1}}\textrm{d}E_R G(E_R,E_O)\phi(E_R)}{\int_{E_{R,j}}^{E_{R,j+1}}\textrm{d}E_{R}\phi(E_R)}.
\end{equation}

When the dimension of the bins is smaller than the energy resolution ({\it i.e.} $G(E_R,E_O)$ does not change significantly within the bin), Eq.~(\ref{eq::defGgen}) can be simplified as
\begin{equation}\label{eq::defGlimit}
    G_{ij}\simeq G(E_{O,i},E_{R,j})\Delta E
\end{equation}
where $\Delta E=E_{O,i+1}-E_{O,i}$. Since this is the case for JUNO, our toy models (and the binning used) will be chosen in such a way that this assumption is verified, and in the next section Eq.~(\ref{eq::defGlimit}) will be used to compute $G_{ij}$.  The result that we will obtain in this section, however, does not depend on the form of $G_{ij}$.

If we have perfect energy resolution, the number of events in a given experiment will follow a Poisson distribution. We have
\begin{equation}\label{eq::defN}
N^R_{H,i}=\overline{N}^R_{H,i} + g_i\sqrt{\overline{N}^R_{H,i}}
\end{equation}
where $g_i$ are Poisson variables shifted and rescaled in such a way that \begin{equation}\label{varG}
\langle g_i\rangle =0 \qquad \langle g_ig_j\rangle = \delta_{ij}.
\end{equation}
Please note that here we are assuming that, in the "real" spectrum, are no intrinsic correlations between bins (as you would have, for example, if you measure the direction of the momenta of the particles created in a two-body decay). 

What would happen with finite energy resolution? 

The number of events in the bin $i$ will be given by the sum of the contributions of events migrating from all the other bins. We have
\begin{equation}\label{defNO}
N^O_{H,i}=\sum_jG_{ij}N^R_{H,j}+m_{ij}\sqrt{N^R_{H,j}}
\end{equation}
where
\begin{equation}\label{rescaleM_multi}
m_{ij}=\frac{M_{ij}-G_{ij}N^R_{H,j}}{\sqrt{N^R_{H,j}}}
\end{equation}
and $M_{ij}$, which follows a multinomial distribution, is the number of events that migrate from bin $j$ to bin $i$. Note that the first two momenta of $m_{ij}$ do not depend on the "real" spectrum, {\it i.e.} on the Gaussian fluctuations there present, indeed
\begin{equation}\label{varM}
\langle m_{ij}\rangle = 0 \qquad 
\langle m_{ij}m_{kl}\rangle = \delta_{jl}(-G_{ij}G_{kl}+\delta_{ik}G_{ij}) .
\end{equation}
This was achieved by construction, due to the shifting and rescaling in Eq.~(\ref{rescaleM_multi}), and it is not true in general, but only for those two momenta.
A consequence of this is that, if we have the average of any number of $g$'s multiplied by one or two $m$'s, they can be factorized:
\begin{equation}\label{varMG}
\langle g_{i1}g_{i2}... m_{ij} \rangle = \langle g_{i1}g_{i2}...\rangle\langle m_{ij} \rangle =0  \qquad 
\langle g_{i1}g_{i2}... m_{ij}m_{kl} \rangle =\langle g_{i1}g_{i2}...\rangle \langle m_{ij}m_{kl} \rangle.
\end{equation}

Using Eq.~(\ref{defNO}) we can write
\begin{equation}
    \delta N_H^i=N_{H,i}^O-\overline{N}_{H,i}^O=\sum_jG_{ij}g^j\sqrt{\overline{N}^R_{H,j}}+\sqrt{N^R_{H,j}(g^j)}m_{ij}.
\end{equation}
In the second term, the notation $N^R_{H,j}(g^j)$ was used to underscore the dependence of $N^R_{H,j}$ on $g^j$.  We then find
\begin{eqnarray}\label{eq::defCorr}
 \textrm{Cov}(N_H^i,N_H^i)&=&\langle (N_{H,i}^O-\overline{N}_{H,i}^O)(N_{H,j}^O-\overline{N}_{H,j}^O) \rangle=\langle\delta N_H^i \delta N_H^j \rangle=
 \nonumber \\
 &&\sum_{k,l}G_{ik}\sqrt{\overline{N}^R_{H,k}}G_{jl}\sqrt{\overline{N}^R_{H,l}}\langle g^k g^l \rangle +\nonumber\\
 &&\sum_{k,l}G_{ik}\sqrt{\overline{N}^R_{H,k}}\langle g^k \sqrt{N^R_{H,l}(g^l)} m_{jl} \rangle + \nonumber \\
 &&\sum_{k,l}G_{jl}\sqrt{\overline{N}^R_{H,l}}\langle \sqrt{N^R_{H,k}(g^k)} g^l m_{ik} \rangle +\nonumber \\
 &&\sum_{k,l}\langle \sqrt{N^R_{H,k}(g^k)} \sqrt{N^R_{H,l}(g^l)} m_{ik} m_{jl} \rangle.
\end{eqnarray}
Due to Eq.~(\ref{varMG}), the second and third terms vanish. The other two are
\begin{equation}\label{eq::varGauss}
    \sum_{k,l}G_{ik}\sqrt{\overline{N}^R_{H,k}}G_{jl}\sqrt{\overline{N}^R_{H,l}}\langle g^k g^l \rangle=\sum_{k}G_{ik}G_{jk}\overline{N}^R_{H,k}
\end{equation}
\begin{eqnarray}
    &&\sum_{k,l}\langle \sqrt{N^R_{H,k}(g^k)} \sqrt{N^R_{H,l}(g^l)} m_{ik} m_{jl} \rangle=\nonumber\\
    %&&\sum_{k,l}\langle \sqrt{N^R_{H,k}(g^k)} \sqrt{N^R_{H,l}(g^l)} \rangle \delta_{kl}(-G_{ik}G_{jl}+\delta_{ij}G_{ik})=\nonumber\\
    %&&\sum_{k}\langle \overline{N}^R_{H,k}+\overline{N}^R_{H,k}g^k \rangle (-G_{ik}G_{jl}+\delta_{ij}G_{ik})=\nonumber\\
    &&\sum_{k} \overline{N}^R_{H,k}(-G_{ik}G_{jl}+\delta_{ij}G_{ik}).
\end{eqnarray}
The correlations between different bins ({\it i.e.} when $i\neq j$) due to the Gaussian and multinomial fluctuation cancel each other and we have
\begin{equation}\label{eq::varJUNO}
 \langle\delta N_H^i \delta N_H^j \rangle= \overline{N}_{H,i}^O\delta_{ij}
\end{equation}
as claimed.

\subsection{Spurious Correlations}

Systematic errors in the model used to fit the data (for example, an error in the energy reconstruction, which would lead to a shift of the observed energy) would lead to an increase of the $\chi^2$; these are sometimes referred to as "correlated errors", however it is worth noticing that, in the statistical sense, no correlation between bins would be present in this case either. 

To see this, consider a new binning, equal to the old binning but shifted by this systematic error.  Then the arguments of the previous subsection run through identically, leading again to Eq.~(\ref{eq::defCorr}) and a lack of correlation between the bins.

%Indeed, such an error would not change the measured number of events in each bin, nor its average \red{Why not?}, which means that the correlation, as defined in Eq.~(\ref{eq::defCorr}), would remain unchanged. 
On the other hand, such a shift would inevitably affect the predicted number of events in each bin, which appears in the expression for $\chi^2$. Indeed, in Eq.~(\ref{eq::defChiGen}), $\overline{N}_{H,i}^O$, {\it i.e.} the expected number of events, must now be replaced by $\tilde{N}_{H,i}^O$, which takes into account the systematic errors in our model. If we define
\begin{equation}
    \Delta \tilde{N}_H^i=\overline{N}_{H,i}^O-\tilde{N}_{H,i}^O
\end{equation}
then Eq.~(\ref{eq::defChi}) would read
\begin{equation}
    \chi^2_N=\sum_i\frac{(\Delta \tilde{N}_{N}^i+\delta N_{N}^i)^2}{\sigma_i^2}, \qquad \chi^2_I=\sum_i\frac{(\Delta \tilde{N}_I^i+\delta N_{N}^i)^2}{\sigma_i^2}
\end{equation}
$\Delta\chi^2$ and $\overline{\Delta\chi^2}$ would now take the form
\begin{equation}
    \Delta\chi^2=\overline{\Delta\chi^2}+\sum_i\frac{2(\Delta \tilde{N}_I^i-\Delta \tilde{N}_N^i)\delta N_{N}^i}{\sigma_i^2}, \qquad \overline{\Delta\chi^2}=\sum_i\frac{(\Delta \tilde{N}_I^i)^2-(\Delta \tilde{N}_N^i)^2}{\sigma_i^2}.
\end{equation}
We notice that these kinds of systematic errors can significantly modify the statistical distribution of $\Delta\chi^2$: first of all, Var($\Delta\chi^2)\neq4\overline{\Delta\chi^2}$; moreover, depending on the values of $\Delta \tilde{N}_I^i$ and $\Delta \tilde{N}_N^i$, $\overline{\Delta\chi^2}$ could even become negative.  This is not surprising, since it is well known that a systematic error on the energy reconstruction, for example, could mimic the opposite mass ordering, leading to the wrong conclusion.

\section{Toy Model(s)}
\label{sec::Toy}
\subsection{Very Simplified Toy Model}

A very simple toy model (VSTM) was considered to study the correlations between bins. The distribution of events with respect to the generic variable $x$ is constant
\begin{equation}
    \frac{dN}{dx}=\textrm{constant}.
\end{equation}
We considered a range of $x$ from 0 to 100, which was divided into 100 bins of size 1. 

Each event is generated separately, then, to simulate the finite resolution of the detector, a Gaussian smearing is applied, with
\begin{equation}
\sigma(x)=5+15\frac{x}{100}   . 
\end{equation}
\begin{figure}[h!]
\centering 
\includegraphics[width=0.5\linewidth]{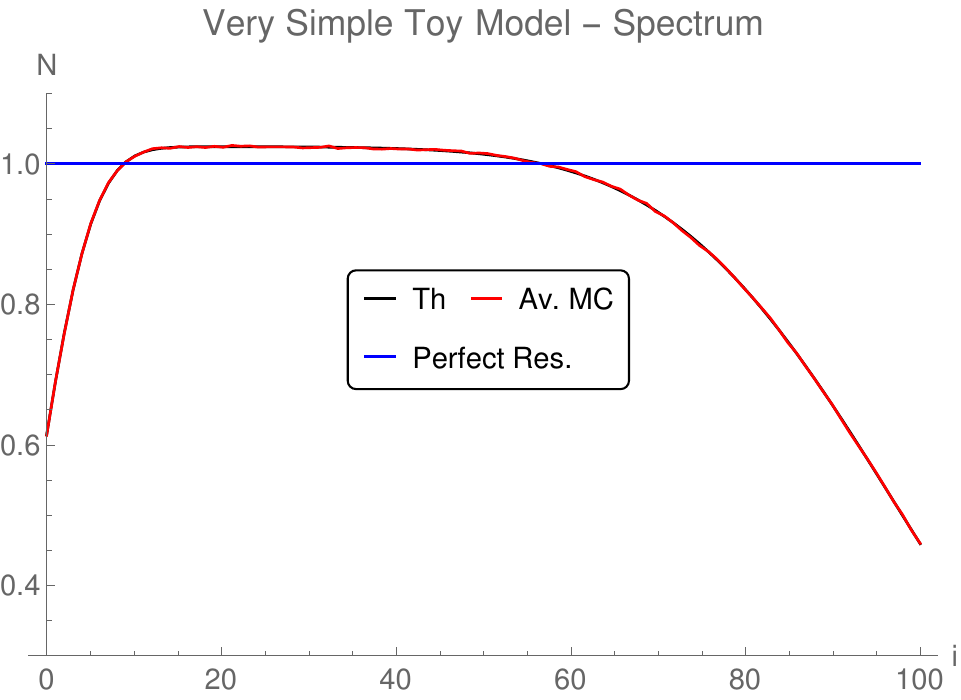}
\caption{\protect\label{fig:SpVSTM} Theoretical spectrum of the VSTM, with perfect and finite resolution. As a sanity check, we also reported the average of the (normalized) spectra obtained in $10^3$ MC simulations (finite resolution), to check that they converge to the theoretical value.}
\end{figure}

Fig. \ref{fig:SpVSTM} reports the spectrum of the model. 

\begin{figure}[h!]
\centering 
\includegraphics[width=0.45\linewidth]{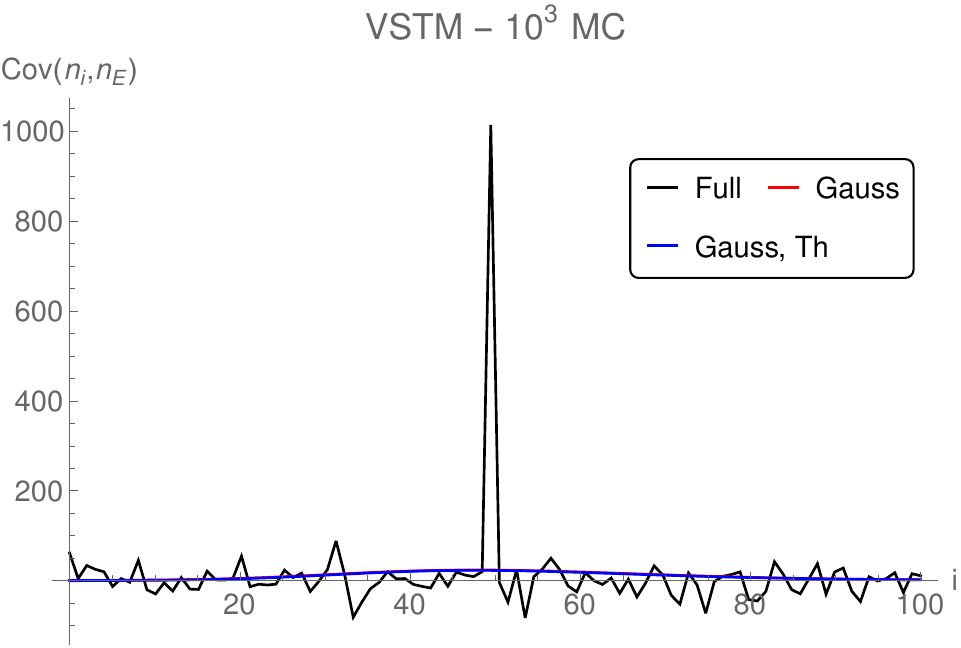}
\includegraphics[width=0.45\linewidth]{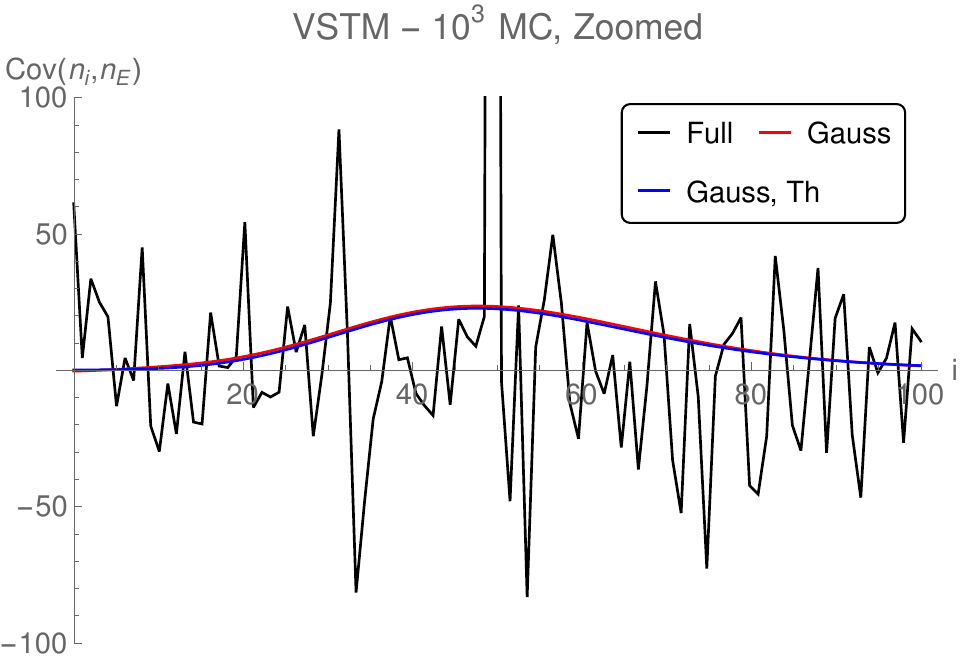}
\caption{\protect\label{fig:CovVSTM-103} Right panel: covariance between the bin with $i=50$ and a generic bin, if all the fluctuations are present or only the Gaussian ones. Left panel: same figure, but zoomed.}
\end{figure}

\begin{figure}[h!]
\centering 
\includegraphics[width=0.45\linewidth]{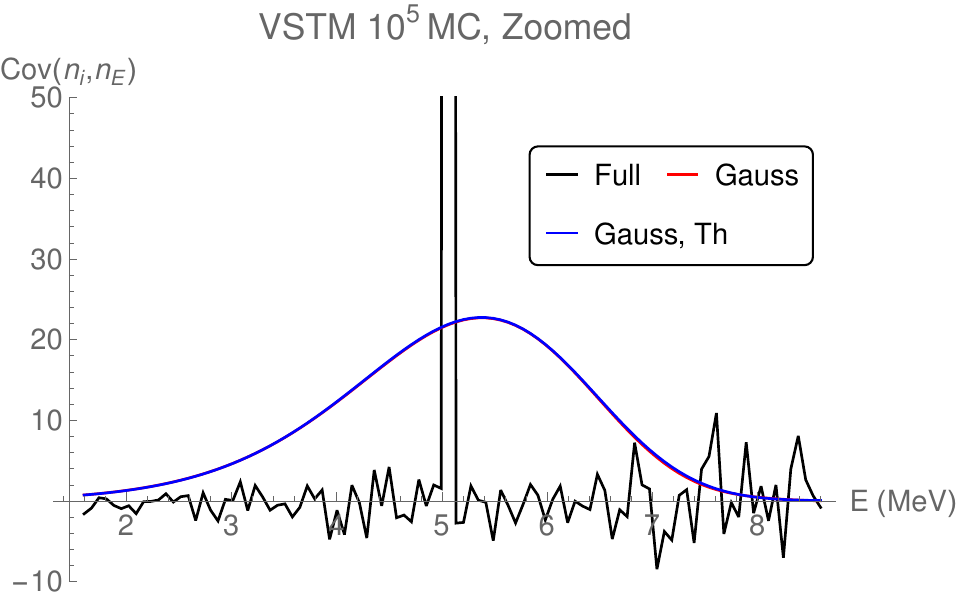}
\includegraphics[width=0.45\linewidth]{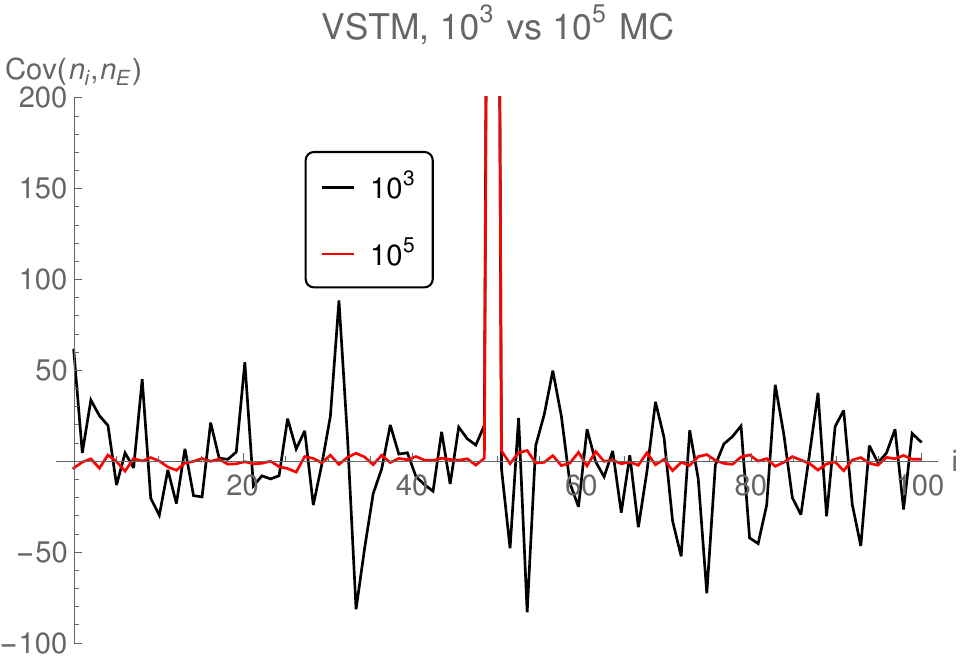}
\caption{\protect\label{fig:CovVSTM-105} Right panel: same as Fig. \ref{fig:CovVSTM-103}, right panel, but with MC sample=$10^5$. Right panel: comparison between MC sample=$10^3$ and MC sample=$10^5$}
\end{figure}

\begin{figure}[h!]
\centering 
\includegraphics[width=0.45\linewidth]{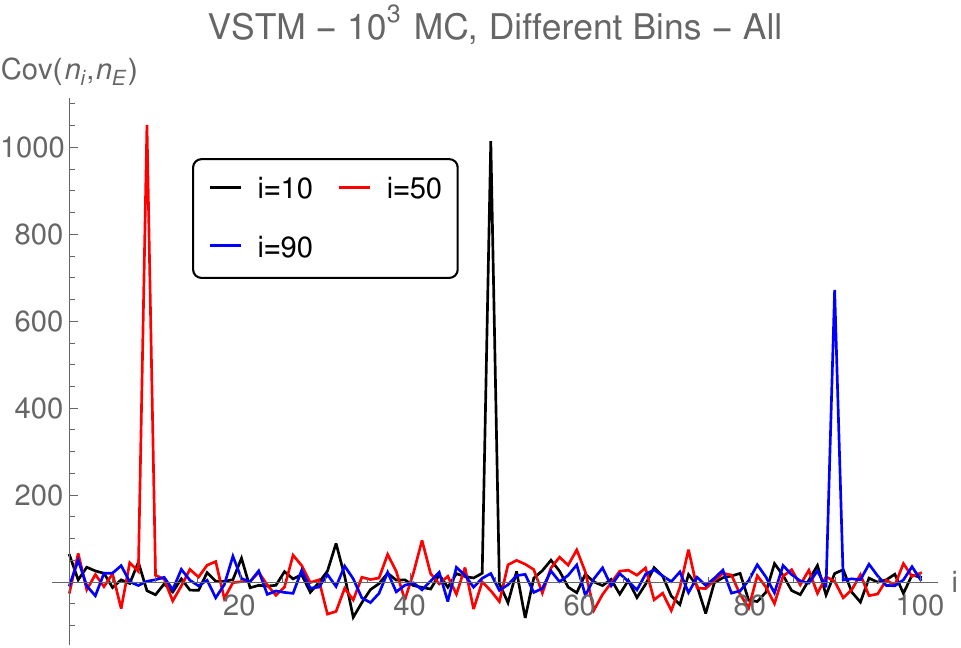}
\includegraphics[width=0.45\linewidth]{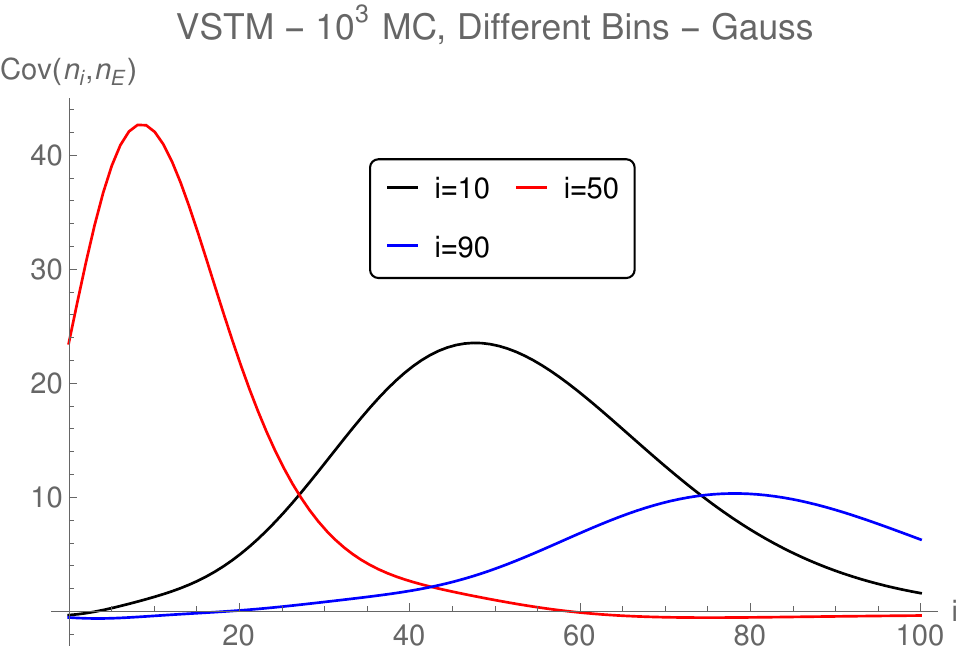}
\caption{\protect\label{fig:CovVSTM-differentBins} Cov($n_i,n_j$), as a function of $j$, when $i=10$, $i=50$, $i=90$. Right panel: all the fluctuations. Left panel: only Gaussian}
\end{figure}

In our simulations, we considered an MC sample of $10^3$ experiments, in each experiment there were $10^5$ events (please note that here and in the rest of the paper the expression ''MC sample" will refer to the former number, not the latter). We considered first the case where all the fluctuations are present.  We then considered also the case where only the Gaussian fluctuations were present. The latter was obtained by summing $\sum_j G_{ij}n_j$, where $n_j$ is the number of events in the bin $j$ obtained in the MC simulations assuming perfect resolution: in this way, the fraction of events migrating from bin $j$ to bin $i$ was always the same, and no multinomial fluctuations are present. As we can see in Fig. \ref{fig:CovVSTM-103} the covariance between bins follows quite closely the expected theoretical behavior: if all the fluctuations are considered, the covariance matrix is diagonal, as can be seen in Eq. (\ref{eq::varJUNO}).  If only the Gaussian fluctuations are taken into account, the covariance follows Eq. (\ref{eq::varGauss}). However, one can notice that, if all the fluctuations are considered, the off-diagonal terms of the covariance matrix are not exactly zero. This is to be expected, but these variations are of the same order of magnitude as the elements of the covariance matrix. To check that they are actually converging to zero, we also considered a MC sample of $10^5$, the results can be found in Fig. \ref{fig:CovVSTM-105}. Fig. \ref{fig:CovVSTM-differentBins} reports Cov($n_i,n_j$), as a function of $j$, when $i=10$, $i=50$, $i=90$ (MC sample=$10^3)$.
\subsection{JUNO Toy Model}
\begin{figure}[h!]
\centering 
\includegraphics[width=0.45\linewidth]{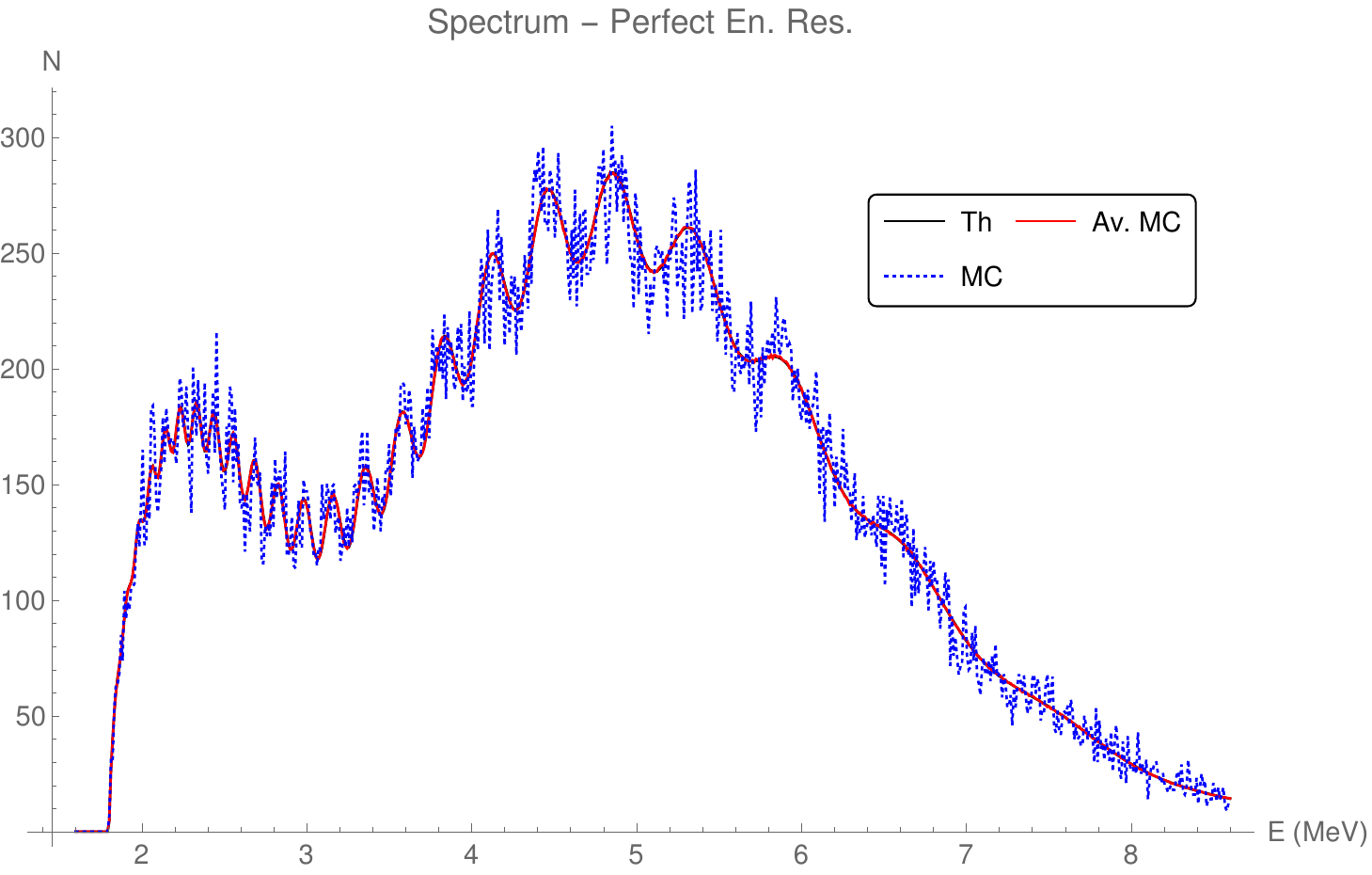}
\includegraphics[width=0.45\linewidth]{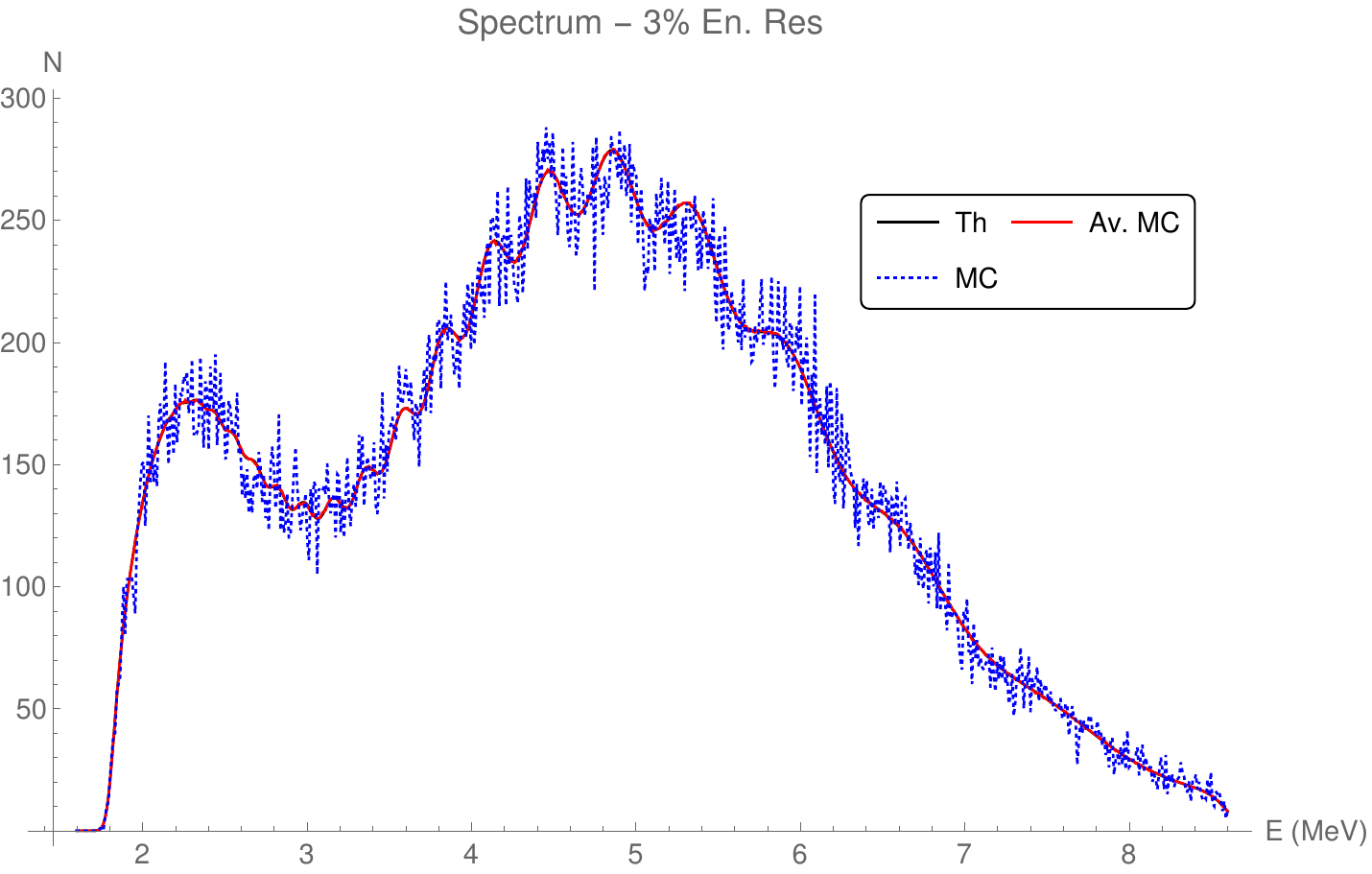}
\caption{\protect\label{fig:JUNOspectrum} Expected spectrum, compared with the average of all the MC sample ($10^3$ experiments) and the spectrum of one single simulation. Left panel: perfect energy resolution. Right panel $\sigma=3\%\sqrt{E}$.}
\end{figure}

\begin{figure}[h!]
\centering 
\includegraphics[width=0.45\linewidth]{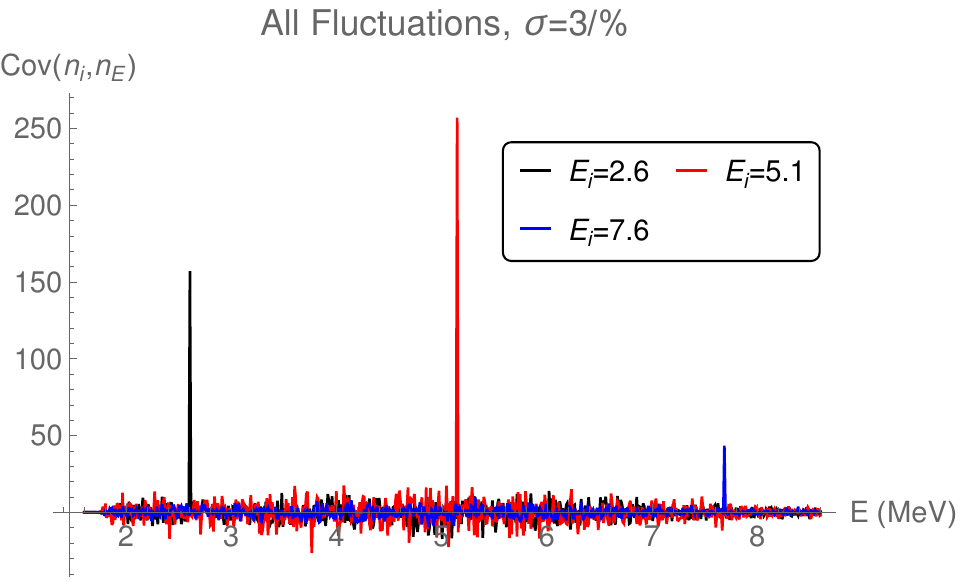}
\includegraphics[width=0.45\linewidth]{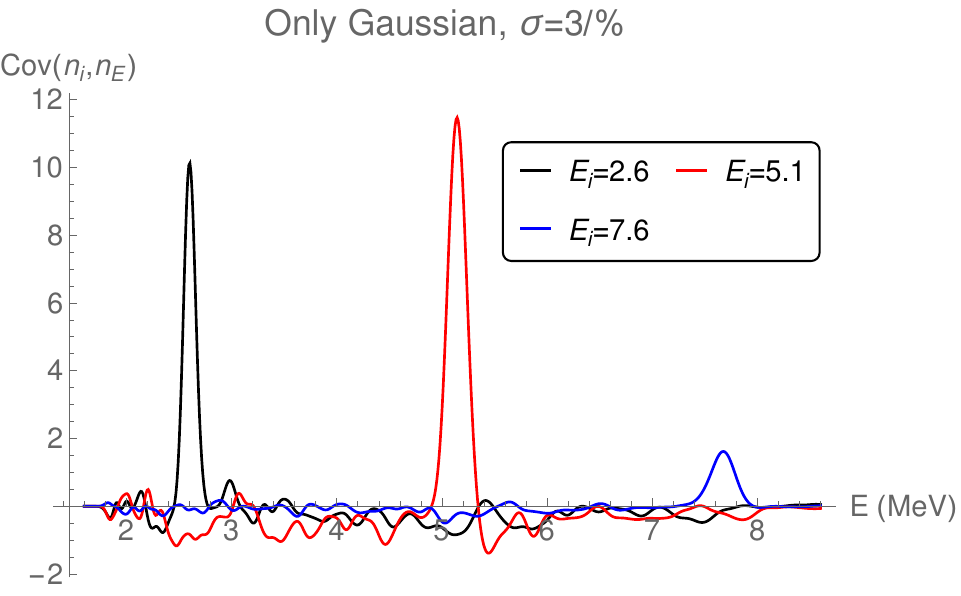}
\caption{\protect\label{fig:JUNO-Cov} Cov($n_i,n_j$), as a function of $j$, when $E_i=2.6$, $E_i=5.1$, $E_i=7.6$ MeV. Right panel: all the fluctuations. Left panel: only Gaussian}
\end{figure}

\begin{figure}[h!]
\centering 
\includegraphics[width=0.5\linewidth]{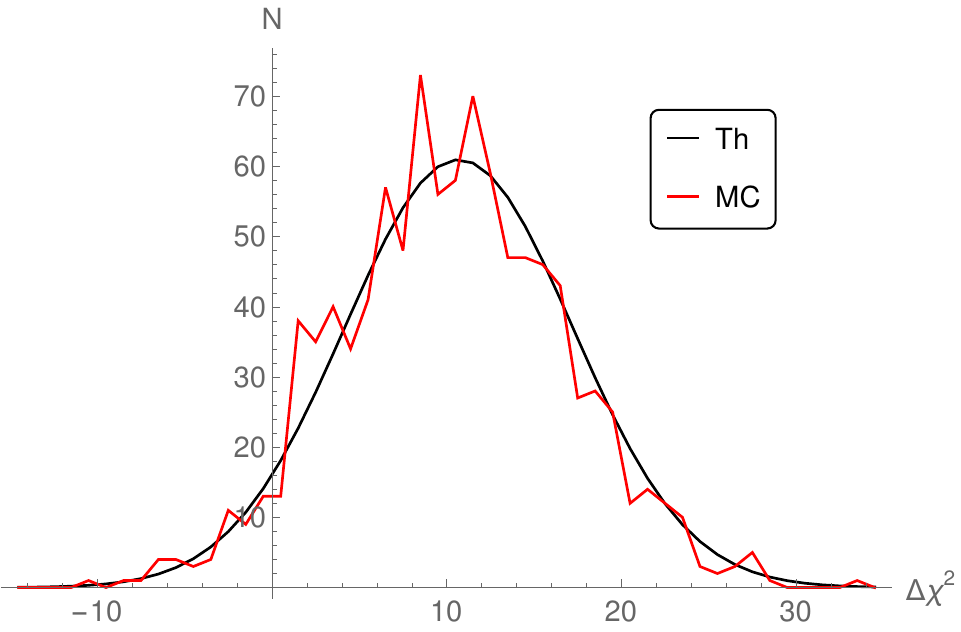}
\caption{\protect\label{fig:HistoChi} $\Delta\chi^2$ from our toy MC simulation. In black, the Gaussian fit. $N_{tot}$ changes, 1 pull parameter considered.}
\end{figure}

We can now check our JUNO toy model. We generated each event separately: the "real" energy was obtained using the expected spectrum at JUNO as a distribution (in Tab. \ref{tab::MixingPar} we report the values of the mixing parameters used).  This is the spectrum that would be seen with perfect energy resolution. For the finite energy resolution case, the "observed" energy was generated, on a event-by-event basis as well, using a Gaussian distribution centered on the "real" energy and with $\sigma=3\%\sqrt{E}$. 
\begin{table}[h!]
    \centering
    \caption{Mixing parameters used in the simulation, taken from \cite{ParticleDataGroup:2022pth}\label{tab::MixingPar}}
    \begin{tabular}{|c|c|c|c|}
\hline
 Sin$^2(\theta_{12})$ & 0.307 &
 Sin$^2(\theta_{13})$ & 2.2$\times10^{-2}$ \\
 \hline
 $\Delta m_{32}^2$ (NH) & ($2.45\pm0.033)\times10^{-3}$ eV$^2$ &
  $\Delta m_{32}^2$ (IH) & $(-2.54\pm0.033)\times10^{-3}$ eV$^2$ 
  \\
  \hline
  $\Delta m_{21}^2$ & 7.53$\times10^{-5}$ eV$^2$ &
  & \\
 \hline
\end{tabular}
\end{table}
In Fig. \ref{fig:JUNOspectrum} we see the expected theoretical spectrum, compared (for a sanity check) with the average of all the experiments (MC sample=$10^3$), and to the spectrum of a single experiment. In Fig. \ref{fig:JUNO-Cov} we report Cov($n_i,n_j$) for different bins, considering all the fluctuations or only the Gaussian ones: we can see that, as in the previous section, it is in good agreement with our theoretical predictions. If all the fluctuations are taken into account, the variation of the off-diagonal terms of the covariance matrix is still pretty large. However, since they are chaotically oscillating, it will not affect the distribution of $\Delta\chi^2$, as we will see soon. Moreover, as shown before, these fluctuations would decrease if the MC sample is increased, so they cannot affect the asymptotical distribution. 
We considered both a Simple vs. Simple scenario, {\it i.e.} no pull parameters were used at all, as well as the case where a single pull parameter was considered, namely $\Delta m_{32}^2$. In Tab. \ref{tab::Results} we report the average and standard deviation of $\Delta\chi^2$ obtained in our toy MC, compared with the theoretical expectation. In the Simple vs. Simple case, we used 3 different values of $|\Delta m^2_{32}|$ when the inverted ordering was considered: $2.54\times10^{-3}$ eV$^2$, which corresponds to the best fit reported in \cite{ParticleDataGroup:2022pth}, $2.57\times10^{-3}$ eV$^2$, which is the value that minimizes the $\chi^2$ computed using the Asimov data set, and $2.43\times10^{-3}$ eV$^2$, which is the value used to generate the data assuming NO. For comparison, in Tab. \ref{tab::Results} we also report the results obtained with the Asimov data set ({\it i.e.} no statistical fluctuations at all). If Fig. \ref{fig:HistoChi} we show the distribution of $\Delta\chi^2$, when 1 pull parameter is considered, compared to the theoretical prediction.
\begin{table}[h!]
    \centering
    \caption{Average and Variance of $\Delta\chi^2$, compared with the expected theoretical value\label{tab::Results}}
    \begin{tabular}{|c|c|c|c|}
\hline
 Case & $\langle\Delta\chi^2\rangle$ & $2\sqrt{\langle\Delta\chi^2\rangle}$ & 
 $\sqrt{\textrm{Var}(\Delta\chi^2)}$  \\
 \hline
 Asimov (1 pull par.) & 10.69 &   6.54 & / \\\hline
1 pull par. & 10.22 & 6.39 & 6.50 \\
 \hline
 0 pull par. $|\Delta m^2_{32}|=2.55\times10^{-3}$& 27.45 & 10.48 & 10.62  \\
 \hline
 0 pull par. $|\Delta m^2_{32}|=2.57\times10^{-3}$& 10.26 & 6.41 & 6.60 \\
 \hline
 0 pull par. $|\Delta m^2_{32}|=2.44\times10^{-3}$& 316.9 & 35.60 & 35.51  \\
 \hline
\end{tabular}
\end{table}

\section{Conclusions}

We have shown that the response of the detector cannot introduce any correlations that are not present in the original spectrum. Indeed, the migration of events between bins caused by the finite resolution would introduce two different kinds of fluctuations: the ones following Poisson distributions, which are present in the original spectrum, and the ones following multinomial distributions, due to the variation in the number of events going from bin $j$ to bin $i$. In the computation of non-diagonal terms of the correlation matrix, these two contributions exactly cancel each other out, leaving only the diagonal term which takes the usual form Var$(N^O_{H,i})=N^O_{H,i}$.

Correlations between bins can nonetheless be present in the original spectrum, however.  For this reason, we have obtained the formula for the variance of $\Delta\chi^2$ in case of correlations between bins when non-nested hypotheses are tested. 

\section* {Acknowledgement}

\noindent
EC thanks Luca Stanco for the useful discussions and suggstions. JE is supported by the NSFC MianShang grants 11875296 and 11675223. JE and EC also thank the Recruitment Program of High-end Foreign Experts for support.

\end{document}